\documentclass[aps,12pt,nofootinbib,amsmath,eqsecnum,floatfix]{revtex4}
\usepackage{graphicx}
\usepackage{amsmath}
\usepackage{amsfonts}
\usepackage{amssymb}
\usepackage{color}
\usepackage{booktabs}
\usepackage{ifpdf}
\usepackage{ulem}

\ifpdf   
  \RequirePackage[pdftex]{hyperref}
\else    
    \RequirePackage[dvips]{hyperref}
  \fi
\hypersetup{bookmarksnumbered,%
              colorlinks,%
               linkcolor=blue,%
               citecolor=blue,%
              plainpages=false,%
            pdfstartview=FitH}

\def\no{\noindent}

\def\bc{\begin{center}}
\def\nno{\nonumber}
\def\ec{\end{center}}
\def\be{\begin{eqnarray}}
\def\ee{\end{eqnarray}}

\newcommand{\omits}[1]{}

\definecolor{dyellow}{rgb}{1.,0.8,.0}
\definecolor{myblue}{rgb}{.1,.1,.7}
\definecolor{dcyan}{rgb}{.0,.6,.6}
\definecolor{dmagenta}{rgb}{0.6,0.0,0.6}
\definecolor{brown}{rgb}{0.6,0.2,0.}
\definecolor{darkblue}{rgb}{.0,.0,0.5}
\definecolor{darkred}{rgb}{0.75,0.0,0.0}
\definecolor{orange}{rgb}{1.,.6,.0}
\definecolor{dorange}{rgb}{0.8,.4,.0}
\definecolor{green}{rgb}{0.0,1.0,0.0}
\definecolor{lightgrey}{rgb}{0.7,0.7,0.7}
\definecolor{purple}{rgb}{.4,.0,.4}

\def\red{\color{red}}

\def\La{\Lambda}
\def\Si{\Sigma}

\def\al{\alpha}
\def\ga{\gamma}
\def\dl{\delta}
\def\eps{\epsilon}

\def\la{\lambda}
\def\th{\theta}
\def\si{\sigma}

\def\d#1#2{\frac{\displaystyle #1}{\displaystyle #2}}
\def\r{\partial}

\newcommand{\dS}{$dS$}
\newcommand{\AdS}{$AdS$}

\newcommand{\vect}[1]{\mbox{\boldmath $#1$}}
\def\P{{\vect P}}
\def\K{{\vect K}}

\def\J{{\vect J}}

\def\NH+{$NH_+$}
\def\NH-{$NH_-$}

\newcommand\btd{\raise 2pt
\hbox{$\hat\bigtriangledown$}\hskip 1.5pt}
\newcommand\bt{\raise 2pt
\hbox{$\bigtriangledown$}\hskip 1.5pt}

\begin{document}



\title{Possible Supersymmetric Kinematics}%
\author{{Chao-Guang Huang}}
\author{{Lin Li}}

\affiliation{
Institute of High Energy Physics and Theoretical Physics Center for
Science Facilities, Chinese Academy of Sciences, Beijing 100049,
China}



\bigskip

\begin{abstract}
The contraction method in different limits to obtain 22 different
realizations of kinematical algebras is applied to study the
supersymmetric extension of \AdS\ algebra and its contractions.  It
is shown that $\frak{p}_2$ $\frak{h}_-$, $\frak{p}'$, $\frak{c}_2$
and $\frak{g}'$ algebras, in addition to $\frak{d}_-$, $\frak{p}$,
$\frak{n}_-$, $\frak{g}$ and $\frak{c}$ algebras, have
supersymmetric extension, while $\frak{n}_{-2}$, $\frak{g}_2$ and
$\frak{g}'_2$ algebras have no supersymmetric extension. The
connections among the superalgebras are established.
\bigskip

\no Keywords: possible supersymmetric kinematics, contraction,
superalgebras

\no

\end{abstract}

\omits{}

\maketitle

\tableofcontents

\section{\ Introduction}

The contraction is a useful method in mathematical physics.  It
reveals the relations among groups and algebras.  It may also be
used to establish the relation among geometries.  By the
In\"on\"u-Wigner (IW) contraction method \cite{IW}, Bacry and
L\'evy-Leblond established the connection among the 11 kinematical
algebras of 8 types \cite{BLL}.  All these algebras satisfy the
assumption that a kinematical group should possess (i) an $SO(3)$
isotropy generated by $\J$, (ii) automorphism of parity
\be \label{parity} %
\Pi: H\to H, \P \to -\P, \K \to -\K, \J \to \J %
\ee
and time-reversal
\be \label{time-reversal} %
\Theta: H\to -H, \P \to \P, \K \to -\K, \J \to \J, %
\ee
and (iii) non-compact one-dimensional
subgroup generated by each boost $K_i$ \cite{BLL}, where $K_i$ is
the components of $\K$.  (The same convention will be used for $\J$
and $\P$, etc.) The 11 kinematical algebras are the Poincar\'e
($\frak{p}$), de Sitter (\dS\ or $\frak{d_+}$), anti-de Sitter
(\AdS\ or $\frak{d}_-$), ihomogenous $SO(4)$ ($\frak{e}'$ or
$\frak{p}'_+$ in literature), para-Poincar\'e ($\frak{p}'$), Galilei
($\frak{g}$), Newton-Hooke ($\frak{n}_+$), anti-Newton-Hooke
($\frak{n}_-$), para-Galilei ($\frak{g}'$), Carroll ($\frak{c}$),
and static ($\frak{s}$) algebras. Relaxing the third condition,
three geometrically kinematical groups --- Euclid ($\frak{e}$),
Riemann ($\frak{r}$), and Lobachevsky ($\frak{l}$) algebras ---
should be added.

The contraction of a kinematical algebra can be studied in two
different ways.  One is just like in Ref. \cite{BLL}.  An algebra is
first defined by a set of the abstract generators.  Then, a
dimensionless parameter $\varepsilon$ is introduced and multiplied
to some generators, which will not alter the algebraic structure.
Finally, the limit of $\varepsilon\to 0$ is taken and the contracted
algebra is attained.  In the manipulation, the realization of the
generators are not used. Hence, it is an abstract way of
contraction.  In the other way, in contrast, an algebra is defined
by a specific realization of a set of generators and the infinite or
zero limit of the parameter(s) in the realization is taken to obtain
the contracted algebra. For example, $\frak{d_\pm}$, realized by a
set of partial differential operators in a given coordinate system,
contract to $\frak{p}$ when the invariant length $l$ tends to
$\infty$ \cite{IPNW}. Similarly, $\frak{p}$ tends to $\frak{g}$ or
$\frak{c}$ when the speed of light $c \to \infty$ or $c\to 0$,
respectively.  This is referred to as the concrete approach of IW
contraction.

In the concrete approach, one may ask:  what is the $l\to 0$ limit
of $\frak{d}_\pm$, is it the same as $\frak{p}$, and does it have
the same physical significance as the ordinary Poincar\'e algebra?
These problems have been studied in Ref.
\cite{{Huang1},{HTWXZ1},{HTWXZ2}}.  It has been shown that when
$\frak{d}_\pm$ are realized in terms of the Beltrami coordinates
\cite{{GHXZ1},{GHXZ2}}, in the $l\to 0$ limit $\frak{d}_{\pm}$ also
contract to an $\frak{iso}(1,3)$ algebra but the generators and thus
the contracted algebra has very different physical significance.
The new realization do not generate the translation invariance of
the Minkowski spacetime.  In fact, the geometry invariant under the
transformations generated by the new realization is a degenerate
one. Therefore, it is called the second realization of Poincar\'e
algebra and denoted by $\frak{p}_2$.  The systematic studies on the
contractions of the Beltrami realization of $\frak{so}(p,5-p)$
($0\leq p \leq 2$) kinematical algebras with two invariant
parameters $c$ and $l$ are made in Ref. \cite{HTWXZ3}.  It has been shown
that there are 22 different realization of possible kinematical
algebras in all, whose generators are all expressed in terms of
coordinate partial differential operators. (The static algebra in
\cite{BLL} has been excluded because its time-translation generator
is expressed in terms of central charge.) All these realizations of
possible kinematics are re-classified and their underlying
geometries are presented.  It is worth mentioning that the 22
realizations of possible kinematical algebras are first obtained by
the combinatorial method in the vector space spanned by the
projective general linear algebra $\frak{pgl}(5,\mathbb{R})$ \cite{GHWZ}.

The possible kinematical superalgebras have also been studied by the
IW contraction method \cite{{Puzalowski},{ClarkLove},{dAG},{HNdO}}.
It has been shown that the Galilei superalgebra can be contracted
from the Poincar\'e superalgebra
\cite{{Puzalowski},{ClarkLove},{dAG},{HNdO}} and that the Poincar\'e
superalgebra can be obtained form \AdS\ superalgebra
$\frak{osp}(1|4)$ \cite{HNdO}.  In Ref. \cite{RembielinskiTybor},
the supersymmetric extension of all possible kinematics in Bacry-L\'evy-Leblond
\cite{BLL} are presented in an abstract way.  In Ref. \cite{CSdT}, the
superalgebras such as supersymmetric extensions of $\frak{p}$, $\frak{g}$, $\frak{c}$,
$\frak{n}_-$, and $\frak{s}$, are re-obtained in the abstract way and
are arranged in a figure to show
their relations.\footnote{The Konopel'chenko algebra
\cite{Konopelchenko} in the figure should be excluded because { there
exists no Hermite representations of generators for the fermionic
part in it \cite{CSdT}}.}  When the concrete realizations
of superaglebras are taken into account, many realizations will be added.
But, in comparison with the contraction scheme in Ref. \cite{HTWXZ3}, not
all the realizations have their supersymmetric extensions.
The purpose of the present paper is to
study the IW contraction of the Beltrami realization of \AdS\
superalgebra and present more realizations of kinematical
superalgebras for completion.

The paper is organized in the following way. In the next
section, we shall briefly review the IW contraction by taking the IW
contraction of Beltrami realization of $\frak{d}_-$ in two opposite
limits, as an example, and list the Beltrami realizations of all
possible kinematical algebras contracting from $\frak{d}_-$. In
section III, the supersymmetric extension of  $\frak{d}_-$ (i.e.
$\frak{osp}(1|4)$) will be reviewed.  In  section IV, we will use the
contraction method to study the supersymmetric extension of the
kinematical algebras in Section II.  The concluding remarks will be
given in the final section.


\section{Beltrami realization of \AdS\ algebra and its contractions}

\subsection{Beltrami realization of \AdS\ algebra}
It is well known that a 4d \AdS\ space-time embedded in
$\mathbb{R}^{2,3}$ with the metric $\eta_{AB}={\rm
diag}(1,-1,-1,-1,1)$,
\be \label{AdS} %
\eta_{AB}\xi^A\xi^B=\eta_{\mu\nu}\xi^\mu \xi^\nu + (\xi^4)^2 = l^2, %
\ee
is mapped onto itself by the group $SO$(2,3), where $A$, $B$,
$\cdots$ run from 0 to 4
and the lowercase Greek letters $\mu$, $\nu$, $\cdots$ 
run from 0 to 3, $\eta_{\mu\nu}={\rm diag}(1,-1,-1,-1)$, and $l$ is
the $AdS$ radius.  The generators for $SO(2,3)$ and the commutation
relations are
\be %
J_{AB} &=& \xi_A\r_{\xi^B}-\xi_B\r_{\xi^A} \\
\, [J_{AB}, J_{CD}] &=& J_{AD} \eta_{BC} - J_{AC} \eta_{BD}
+ J_{BC} \eta_{AD} - J_{BD}\eta_{AC}.  \label{am-c} %
\ee
In terms of the 4d Beltrami coordinates
\be \label{BeltramiCoordinates} %
 x^\mu=l \d{\xi^\mu}{\xi^4}, %
\ee
the generators read
\be \begin{cases} %
(l P_\mu^-):=J_{4 \mu}  = l(\r_{x^\mu}+l^{-2}x_\mu x^\nu\r_{x^\nu}) &\\
J_{\mu\nu}= x_\mu \r_{x^\nu}-x_\nu \r_{x^\mu}. &
\end{cases} \label{generatorinBeltrami} %
\ee
In 3d realization, they are
\be \begin{cases} %
H^-=\r_t+\nu^2 t x^\mu \r_{x^\mu}, \qquad {P}_i^-=\r_i+l^{-2}x_ix^{\mu}\r_{x^\mu},\\
K_i=t\r_i-c^{-2}x_i \r_t,\qquad \ \ J_{ij}=x_i\r_j-x_j\r_i \quad
\mbox{or} \quad J_i=\d 1 2 \eps_{i}^{\ jk} (x_j\r_k-x_k\r_j),
\end{cases}\label{3dgeneratorBeltrami} %
\ee
where $H^-=cP_0^-$, $t=x^0/c$, $\nu=c/l$, $\r_t=\frac \r {\r
t}$, $\r_i=\frac \r {\r x^i}$, $x_i=-\dl_{ij}x^j$.  The dimensions
of $H^-$, $P_i^-$, and $K_i$ are $T^{-1}$, $L^{-1}$, and $c^{-1}$,
respectively.  \omits{Eqs. (\ref{generatorinBeltrami}) and
(\ref{3dgeneratorBeltrami}) are the 4d and 3d Beltrami realizations
of $\frak{d}_-$, respectively.}

\subsection{IW contraction of Beltrami realization of $AdS$ algebra
in two opposite limits}

Suppose the generators $T_I\ (I=1,\cdots n)$ span the Lie algebra of
a Lie group
\be %
[T_I, T_J]=C_{IJ}^K T_K . %
\ee
Under a linear homogeneous non-singular transformation,
\be \label{transform} %
S_I=U^{\ J}_{I}T_J , %
\ee
the structure of the algebra will not
change though the structure constants $C_{IJ}^K$ will be replaced by other
constants.  If the first $m<n$ generators span a subalgebra of the
algebra and the matrix $U$ in Eq.(\ref{transform}) takes the form
\be \label{TransformMatrix} %
(U^{\ J}_{I})=\left (\begin{array}{cc} I_{m} & 0 \\
0 & \varepsilon I_{n-m}\end{array} \right ), %
\ee
where $I_m$ and $I_{n-m}$ are unit matrices, then the transformation
(\ref{transform}) becomes singular when $\eps \to 0$ and will lead
to a new algebra \cite{IW}. The operation is known as the IW
contraction with respect to the subalgebra.

In order to study the explicit contraction of the Beltrami
realization of $\frak{d}_-$, the invariant parameters $l$ and/or $c$
in Eqs. (\ref{BeltramiCoordinates})--(\ref{3dgeneratorBeltrami}) are
replaced by running parameter $l_r$ and/or $c_r$, respectively, which are still
invariant under the \AdS\ transformations.  As the result, all
quantities in Eqs.
(\ref{BeltramiCoordinates})--(\ref{3dgeneratorBeltrami}) will be
replaced by the running ones.   Then, multiplying suitable parameter
in terms of $l_r$ and/or $c_r$ and taking the limit, one may obtain
the contracted algebras.

For example, the Beltrami realization of $\frak{d}_-$ can be
contracted explicitly in the following two ways to obtain the
completely different realizations of the Poincar\'e algebra.    At
any point in the neighborhood of the ``north pole" of Eq.
(\ref{AdS}), we have
\be %
\xi^\mu \approx 0,\qquad  \xi^4 \approx l_r. %
\ee
Following Eqs. (\ref{transform}) and (\ref{TransformMatrix}), the
generators of $SO(2,3)$ can be written as \cite{Gilmore}
\be \label{lr2infty} %
\left (\begin{array}{c}J_{\mu\nu}\\ P^-_\mu\end{array}\right) =\left
(\begin{array}{cc}I_6 & 0 \\ 0 & \varepsilon I_4\end{array}\right)
\left(\begin{array}{c}J_{\mu\nu}\\J_{4 \mu }\end{array}\right)
\qquad \qquad \mbox{with } \varepsilon=l_r^{-1}. %
\ee
In the limit $l_r\to \infty$ or $\varepsilon \to 0$, we have
\be %
&&J_{\mu\nu}=x_\mu\r_{x^\nu}-x_\nu\r_{x^\mu} \to J_{\mu\nu}\\
&&P^-_\mu=\d {1}{l_r}J_{4\mu} =\d 1 {l_r}
(\xi_4\r_{\xi^\mu}-\xi_\mu\r_{\xi^4}) = (\r_{x^\mu}+ l_r^{-2}x_\mu
x^\nu\r_{x^\nu}) \to  \r_{x^\mu}=:P_\mu. \ee
They are the generators of the ordinary Poincar\'e group. 

On the other hand, one may use an alternative set of generators
\be\label{lr20} %
\left (\begin{array}{c}J_{\mu\nu}\\
\Pi^-_\mu\end{array}\right) =\left ( \begin{array}{cc}I_6 & 0 \\ 0 &
\varepsilon' I_4\end{array}\right) \left(\begin{array}{c}J_{\mu \nu
} \\ J_{4 \mu
}\end{array}\right)\qquad \qquad \mbox{with } \varepsilon'= l_r/l^2. %
\ee
In the limit $l_r \to 0$,
\be %
&&J_{\mu\nu}=x_\mu\r_{x^\nu}-x_\nu\r_{x^\mu} \to J_{\mu\nu} \\
&&\Pi^-_\mu= \d {l_r}{l^2}J_{4\mu } = \d {l_r}{l^2}
(\xi_4\r_{\xi^\mu}-\xi_\mu\r_{\xi^4})= \d {l_r^2}{l^2}(\r_{x^\mu}+
l_r^{-2}x_\mu x^\nu\r_{x^\nu})\to l^{-2}x_\mu
x^\nu\r_{x^\nu}=:P'_\mu.\qquad \quad %
\ee
They also span an
$\mathfrak{iso}(1,3)$ algebra. However, $P'_\mu$ are obviously
different from $P_\mu$. The new realization of $\frak{iso}(1,3)$ do
not generate the translations on a Minkowski space-time, and thus
$P'_\mu$ are called the pseudo-translation generators.  (It should
be noted that the definitions of the pseudo-translation generators
here are different by a minus from the ones in
\cite{{GHWZ},{Huang1},{HTWXZ1},{HTWXZ2},{HTWXZ3}}, which do not
affect the algebraic structure.) For brevity, the new realization of
$\frak{iso}(1,3)$ is referred to as the second Poincar\'e algebra
($\frak{p}_2$).  The same nomenclature also applies to the other
kinematical algebras.

Obviously, the ordinary and the second Poincar\'e algebras are the
contraction of the same $AdS$ algebra in two opposite limits.
Algebraically, the ordinary and the second Poincar\'e algebras are
identical to each other.  If the abstract generators are dealt with
just like the treatment in Ref. \cite{BLL}, the second Poincar\'e
algebra cannot be distinguished from the ordinary one.  However, if
the realization is taken into account, the ordinary and second
Poincar\'e algebras are dramatically different from each other in
physics and geometry. For example, the underlying geometries for the
second Poincar\'e algebra must be degenerate ones rather than
a 4d Minkowski space-time
\cite{{Huang1},{HTWXZ1},{HTWXZ2}}. \omits{\red (In fact, one may get
more Poincar\'e algebras starting from the different realization of
$AdS$ algebra \cite{AP}.)}

\begin{table}[th]
\caption{\quad \AdS\ algebra and its contractions} {\small
\begin{tabular}{|c c c c c c c c c|}
\hline Algebra & Symbol & Generator set
  & $[\cal H,\vect{\cal P}]$ \quad & $[\cal H,\vect{\cal K}]$ \quad &
  $[\vect{\cal P},\vect{\cal P}]$\quad
  &$[\vect{\cal K},\vect{\cal K}]$ \quad &$[\vect{\cal P},\vect{\cal K}]$ & Limit\\
\hline  \AdS & $\mathfrak{d}_-$ & $(H^-, \P^-, \K, \J)$ & $-\nu^2\K
$ & $ \P^-$ & $-l^{-2}\J $
& $-c^{-2}\J$ & $c^{-2} H^-$ & \\
{\it Poincar\'e} &$\begin{array}{c}\mathfrak{p}\\
\mathfrak{p}_2\end{array}$ & $\begin{array}{c}(H, \P, \K, \J)\\
(H', \P', \K, \J)\end{array}$ & 0 & $\begin{array}{c}\P \\
\P'\end{array}$ & 0 &$-c^{-2}\J$&$\begin{array}{c}c^{-2} H \\ c^{-2}
H'\end{array}$ &
$\begin{array}{c}l_r\to\infty \\ l_r\to 0 \end{array}$\\
\hline
{\it Galilei}&$\begin{array}{c}\mathfrak{g}\\
\mathfrak{g}_2\end{array}$&$\begin{array}{c}(H, \P, \K^{\frak g},\J)\\
(H', \P', \K^{\frak c},\J)\end{array}$&0 & $\begin{array}{c}\P\\
\P'\end{array}$
&  0 &   0 &   0  &  $\begin{array}{c}l_r, c_r\to\infty, \nu_r\to 0 \\
l_r, c_r\to 0, \nu_r \to \infty \end{array}$\\
{\it Carroll}&$\begin{array}{c}\mathfrak{c}\\
\mathfrak{c}_2\end{array}$ & $\begin{array}{c}(H, \P, \K^{\frak c},\J )\\
(H', \P', \K^{\frak g},\J )\end{array}$ & 0& 0& 0& 0&$\begin{array}{c}c^{-2} H \\
c^{-2} H'\end{array}$ & $\begin{array}{c}l_r\to\infty, c_r\to 0 \\
l_r\to 0, c_r\to \infty \end{array}$\\
${NH}_-$  &$\begin{array}{c}\mathfrak{n}_-\\
\mathfrak{n}_{-2}\end{array}$ & $\begin{array}{c}(H^-, \P, \K^{\frak g},\J )\\
(H^-, \P', \K^{\frak c},\J )\end{array}$ & $\begin{array}{c}-\nu^2\K^{\frak g}\\
-\nu^2\K^{\frak c}\end{array}$ &$\begin{array}{c}\P\\
\P'\end{array}$ &0 &0 &0 &
$\begin{array}{c}l_r, c_r\to\infty, \nu_r=\nu \\
l_r, c_r\to 0, \nu_r=\nu \end{array}$\\
{\it para-Galilei}&$\begin{array}{c}\mathfrak{g}'\\
\mathfrak{g}'_2\end{array}$ & $\begin{array}{c}(H', \P, \K^{\frak g}, \J)\\
(H, \P', \K^{\frak c},\J )\end{array}$ & $\begin{array}{c}-\nu^2\K^{\frak g}\\
-\nu^2\K^{\frak c}\end{array}$ & 0 & 0 & 0 &0 &
$\begin{array}{c}l_r, c_r, \nu_r\to\infty \\ l_r,c_r,\nu_r\to 0 \end{array}$\\
$HN_-$&$\begin{array}{c}\mathfrak{h_-}\\
\mathfrak{p}'\end{array}$&$\begin{array}{c}(H, \P^-, \K^{\frak c},\J)\\
(H', \P^-, \K^{\frak g},\J )\end{array}$&$\begin{array}{c}-\nu^2\K^{\frak c}\\
-\nu^2\K^{\frak g}\end{array}$ &0 &$-l^{-2}\J$&0 &$\begin{array}{c}c^{-2} H \\
c^{-2} H'\end{array}$ &$\begin{array}{c}c_r\to 0 \\ c_r\to \infty \end{array}$\\
\hline
\end{tabular}
}
\end{table}

With the same technique, one may obtain other 10 realizations of
contracted algebras, which have the same $\frak{so}(3)$ spanned by
\J\ \cite{HTWXZ3,GHWZ}. TABLE I lists the generators and commutation
relations of $\frak{d}_-$ and its contractions, except the common
commutation relations involving \J.  In TABLE I, the generators are
defined by
\be %
\begin{cases} H=\r_t,\quad H'=\nu^2tx^\mu\r_{x^\mu},\\
P_i=\r_i,\quad P'_i=l^{-2}x_i x^\mu \r_{x^\mu}, \\
{K}_i^{\frak{g}}=t\r_i,\quad {K}_i^{\frak{c}}=-c^{-2}x_i \r_t,\\
\end{cases}
\ee
in addition to the definition of Eq. (\ref{3dgeneratorBeltrami}),
and ${\cal H}$, $\vect{\cal P}$, or $\vect{\cal K}$ stands for the
suitable one in $\{H^-, H, H'\}$, $\{\vect{P}^-, \vect{P},
\vect{P}'\}$, or $\{\vect{K}, \vect{K}^{\frak g}, \vect{K}^{\frak
c}\}$, respectively. The name of anti-Hooke-Newton algebra $HN_-$
comes form that it is different from the anti-Newton-Hooke algebra
$NH_-$ by the replacement $H^-\leftrightarrow H$, $P_i
\leftrightarrow P_i^-$, $K_i^{\frak g}\leftrightarrow K_i^{\frak c}$
\cite{GHWZ}. It is also called the para-Poincar\'e algebra
\cite{BLL}. From the geometrical point of view, the first version of
$HN_-$ algebra is referred to as the anti-Hooke-Newton algebra
($\frak{h}_-$),  while its second version is para-Poincar\'e algebra
($\frak{p}'$) \cite{HTWXZ3}.  In TABLE I, the static algebra has
been excluded because its time-translation generator $H^{\frak s}$
is meaningful only when the central extension is considered.  It
should be remarked that in comparison with the table in Ref.
\cite{GHWZ} and \cite{HTWXZ3} the minus in the definitions of the
pseudo-translation generators $H'$ and $\vect{P}'$ have been
removed, thus the sets of generators for $\frak{n}_{-2}$ and
$\frak{p}'$ are modified correspondingly, and the structure
constants of $\frak{g}'$ have an overall minus, which do not affect
the algebraic structure.


\section{Supersymmetric extension of $AdS$ algebra}

The supersymmetric extension of $\frak{d}_-$ can be established on a
superspace spanned by coordinates $\xi^A$ of dimension $L$ subject
to Eq. ({\ref{AdS}}), a Majorana fermionic coordinate $\th$ of
dimension $L^{1/2}$. Concretely, we  may choose
the Weyl basis of the Dirac matrices
\be \ga^\mu=\left(\begin{array}{cc} 0 & \si^\mu\\ \bar \si^\mu & 0
\end{array} \right), \qquad
\ga^5 =i\ga^0\ga^1\ga^2\ga^3=\left(\begin{array}{cc} -\si^0 & 0\\
0 & \bar \si^0 \end{array} \right) =:\ga^4, \ee
and take the charge conjugation matrix
\be \label{C}%
C = i\ga^2\ga^0=i\left (\begin{array}{cc}\si^2\bar\si^0 &0\\
0& \bar \si^2 \si^0 \end{array} \right ), \ee
satisfying $C\ga^\mu C^{-1}=-(\ga^\mu)^T$, where the superscript $T$
denotes the transpose,
\be %
\si^\mu=(\si^0, \si^i)=(I_2, \tau^i) \quad \mbox{and} \quad
\bar \si^\mu=(\bar\si^0, \bar \si^i) =(I_2, -\tau^i).%
\ee in which $I_2$ is the $2\times 2$ unit matrix and $\tau^i$ are 3
Pauli matrices. Acting on a scalar superfield, the bosonic generators
are extended as
\be \label{sads-bg1} %
J_{\mu\nu} = x_\mu\r_\nu -x_\nu\r_\mu +
\bar\th\Si_{\mu\nu}\frac \r {\r \bar\th}
\ee
and
\be \label{sads-bg2} %
P_\mu^-=\frac \r {\r x^\mu}+\frac 1 {l^2} x_\mu x^\nu \r_\nu+\frac 1 {l^2}
 \frac {x^\nu} {1+\sqrt{\si}} \bar\th \Si_{\nu\mu}\frac \r {\r \bar\th},%
\ee
where $\bar \th$ is the Dirac conjugate of $\th$,
\be \label{Si}%
\Si^{\mu\nu}=\d 1 4[\ga^\mu, \ga^\nu], %
\ee
and $\si(x):=1+l^{-2}\eta_{\mu\nu}x^\mu x^\nu>0 $
defines the domain of Beltrami-\AdS\ spacetime.
The fermionic generator $ Q $ and its conjugate $\bar Q$, which
obey the Majorana condition
 $ Q=C\bar Q ^T $ and are of dimension $L^{-1/2}$, may be chosen as \cite{IvanovSorin}
\begin{eqnarray} \label{sads-fg1} %
 \nonumber Q &=& (1-\frac 1 {4l}\bar\th \th)\La \left \{i(1+\frac 1 {4l} \bar\th \th)\frac
\r {\r \bar\th}+\frac i l \th\bar\th \frac \r {\r \bar \th}
-\frac i {4l}(\ga^\nu \th)\bar\th\left (\ga_\nu-\frac {2i} l \frac {x^\mu}{1+\sqrt
\si}\Si_{\mu\nu}\right )\frac{\r}{\r \bar\th}\right . \\
& &\left . -\frac 1 2 \sqrt{\si}\ga^\nu\th \left (\frac \r {\r x^\nu }
+\frac {l^{-2}x_\nu x^\mu}{1+\sqrt{\si}} \r_\mu \right )\right \} ,
\end{eqnarray}
\begin{eqnarray} \label{sads-fg2} %
 \nonumber \bar Q &=& - (1-\frac 1 {4l}\bar\th \th)\left \{i(1+\frac 1 {4l} \bar\th \th)\frac
\r {\r \th}\La^{-1} -\frac i l (\bar \th\La^{-1})\bar\th \frac \r {\r \bar \th}
-\frac i {4l}(\bar \th \ga^\nu \La^{-1})\bar\th\left (\ga_\nu-\frac {2i} l \frac {x^\mu}{1+\sqrt
\si}\Si_{\mu\nu}\right )\frac{\r}{\r \bar\th}\right .\\
& &\left .-\frac 1 2 \sqrt{\si}(\bar \th\ga^\nu \La^{-1})\left (\frac \r {\r x^\nu }
+\frac {l^{-2}x_\nu x^\mu}{1+\sqrt{\si}} \r_\mu\right )\right \} ,
\end{eqnarray}
where
\be  %
\La=\left (\frac {1+\sqrt \si}{2\sqrt \si}\right )^{\frac 1 2}\left (I+ \frac i l \frac {x^\mu}
{1+\sqrt \si}\ga_\mu\right ), \qquad I \mbox{ is a }4\times 4 \mbox{ unit matrix.}%
\ee

The commutators among the pure bosonic generators $J_{\mu\nu}$ and
$P^-_\mu =l^{-1}J_{4\mu}$ with the supersymmetric extensions (\ref{sads-bg1}) and
(\ref{sads-bg2}) are
still given by Eq.(\ref{am-c}).   The structure relations of $\frak{osp}(1|4)$
involving fermionic parts are
\be
   \left[J_{ij},Q\right]=-\Si_{ij}Q, \ \quad && \quad \left[K_i,Q\right]=   - \d {1}{c} \Si_{0i}Q,
\label{J^'Q}\\
\left[H^-,Q\right]=\d {i\nu} {2} \ga_0 Q
  , \quad && \quad
 \left[P_i^-,Q\right]=\d {i} {2l}\ga_{i}Q , \label{PQ}%
\ee
\be
 \{Q,\bar{Q}\}=\frac i c \ga^0 H^- +i\ga^iP^{-}_i-\frac 1 l
\Si^{ij}J_{ij}-\frac{2c} l \Si^{0i}K_i.
 \label{QbarQbar}
\ee

When the invariant parameters $l$ and/or $c$ vary finitely and are labeled
by $l_r$ and/or $c_r$, respectively, the  \AdS\ superalgebra spanned by $(H_r^{-},
\vect{P}_r^{-}, \vect{K}_r, \vect{J}, Q_r, \bar Q_r)$ will remain
the form of Eqs. (\ref{am-c}),   (\ref{J^'Q}),
(\ref{PQ}), and (\ref{QbarQbar}) with $l$ and/or $c$ replaced by $l_r$ and/or $c_r$.  That is,
\be %
\, [J_i, J_j] =-\eps_{ij}{}^kJ_{k}, \qquad \ \ &&
\, [J_{i}, (K_r)_{j}] =  - \eps_{ij}{}^{k}(K_r)_{k},  \\   %
\, [J_{i}, H^-_r] = 0,\qquad \qquad \ \ \,  &&
\, [J_{i}, (P^-_r)_{j}] =  - \eps_{ij}{}^{k}(P^-_r)_{k},  \\ %
\left[J_{i},Q_r\right]=-\frac 1 2 \eps_i{}^{jk}\Si_{jk}Q_r,  && \,
\left[(K_r)_i,Q_r\right]= - \d {1}{c_r} \Si_{0i} Q_r,
\label{JQ}\\
\left[H_r^-,Q_r\right]=\d {ic_r} {2l_r} \ga_{0 } Q_r,  \quad \ && \,
 \left[(P_r^-)_i,Q_r\right]=\d {i} {2l_r}\ga_{i}Q_r, 
\ee
and
\be %
\{Q_r,\bar{Q}_r\}=\frac i {c_r} \ga^0 H_r^- +i\ga^i (P_r^-)_i -\frac 1
l_r \Si^{ij}J_{ij}-\frac{2c_r} {l_r} \Si^{0i} (K_r)_i. \label{QbarQbar-r}
\ee

Further, the \AdS\ superalgebra still remains if $\th$ and $\bar \th$ undergo
a scale transformation
\be \label{th-zeta2} %
\th \to {\epsilon} \th, \qquad \bar \th \to {\epsilon}\bar \th  %
\ee
in addition to the replacement of $l$ and/or $c$ by $l_r$ and/or $c_r$, respectively.

\bigskip


\section{Possible Kinematical Superalgebras}

The contraction of the Beltrami realization of \AdS\ superalgebra in
different limits can be attained in the same way as the
contraction of the Beltrami realization of \AdS\ algebra.  In all contraction,
the supersymmetric extensin of $J_i=\frac 1 2 \eps_i{}^{jk}J_{jk}$ remains unchanged:
\be
J_{ij} = x_i \r_j -x_j\r_i + \bar \th \Si_{ij} \frac \r {\r \bar \th}. \label{SAM}
\ee
In the
following, we shall consider these contractions one by one.

\subsection{Two realizations of the Poincar\'e superalgebra}

Recall that the ordinary and second realization of the  generators
of Poincar\'{e} algebra can be contracted from the generators of
$AdS$ algebra in the following way,
\be %
 {\frak p}: &&H=\lim_{l_r\to \infty}H_r^-,\qquad  \vect{P}=\lim_{l_r\to
 \infty}\vect{P}_r^-,\quad \ \vect{K},\quad \ \vect{J},\\
{\frak p}_{2}:&&
 H'=\lim_{l_r\to 0}\d{l_r^2}{l^2}H_r^-,\quad \, \vect{P}'=\lim_{l_r\to
 0} \d{l_r^2}{l^2}\vect{P}_r^-,\ \ \vect{K}, \quad \ \vect{J}. %
\ee
After supersymmetric extension and in the same limits,
\be
 {\frak p}: &&H=\r_t,\qquad  \qquad P_i=\r_{x^i},\qquad \qquad  K_i=t\r_{x^i}-\frac 1 {c^2}x_i\r_t+
\frac 1 c \bar \th \Si_{0i}\frac \r {\r\bar \th},\\
{\frak p}_{2}:&&
 H'=\frac {c^2}{l^2}tx^\mu\r_{x^\mu},\ \, P'_i=l^{-2}x_ix^\mu\r_{x^\mu},\ \
 K_i=t\r_{x^i}-\frac 1 {c^2}x_i\r_t+
\frac 1 c \bar \th \Si_{0i}\frac \r {\r\bar \th},
\ee
plus Eq.(\ref{SAM}).
Their commutators are pure bosonic parts for the corresponding contracted superalgebras.

In the limit of $l_r \to \infty$, the fermionic generator reduces to
that of Poincar\'e superalgebra, as expected,
\be %
Q^{\frak{p}} :=\lim_{l_r\to\infty}Q_r= i \frac \r {\r
\bar\th}-\frac 1 2 \ga^\mu \th \frac \r {\r x^\mu} %
\ee
and the (anti-)commutators of super \AdS\ algebra,
(\ref{J^'Q}), (\ref{PQ}) and (\ref{QbarQbar}), contract to those for
Poincar\'{e} superalgebra:
\be  %
\, [J_{ij}, Q^{\frak{p}}]=-\Si_{ij}Q^{\frak{p}},
&&  [K_{i}, Q^{\frak{p}}]=-\frac 1 c\Si_{0i} Q^{\frak{p}},\\
\, [H, Q^{\frak{p}}] =0, \qquad  \quad &&
 [P_i, Q^{\frak{p}}]=0, \\
\{Q^{\frak{p}}, \bar{Q}^{\frak{p}}\} = i\ga^\mu P_\mu. &&
   \label{P0inPoin}
\ee

On the other hand, as $l_r \to 0$ the $AdS$ superalgebra with scaled $\th$ also
contracts to the Poincar\'{e}
superalgebra if $\epsilon \to 0$ according to $l_r/l$, denoted by $\varepsilon_m^2$,
and the fermionic generator is chosen as
\be %
Q^{\frak{p}_2}
&:=&\lim_{\substack{l_r\to  0}}\varepsilon_m^2 Q_r
=\La^{\frak{p}_2}\left (i\frac \r{\r\bar\th}-\frac 1 {2l^2} (\ga^\nu \th) x_\nu
 x^\mu \r_\mu \right ), \\
\mbox{with\qquad }\La^{\frak{p}_2}&:=&\lim_{\substack{l_r\to  0}}\La_r=
\frac 1 {\sqrt 2}\left (I+i \frac {x^\mu \ga_\mu}{\sqrt
{x\cdot x}}\right ) \\
\mbox{and\qquad }x\cdot x &=&\eta_{\mu\nu}x^\mu x^\nu. \notag%
\ee
The fermionic parts of algebraic relations read
\be %
\left[J_{ij},Q^{\frak{p}_2}\right] = -\Si_{ij} Q^{\frak{p}_2}, &\quad &
\left[K_{i},Q^{\frak{p}_2}\right]
  =-\frac 1 c \Si_{0i} Q^{\frak{p}_2}, \\
\ [H', Q^{\frak{p}_2}]=0, \qquad \quad \,  &\quad &
 [P'_{i}, Q^{\frak{p}_2}]=0,  \\
\{Q^{\frak{p}_2}, \bar Q^{\frak{p}_2}\}=i\ga^\mu P'_\mu , \ &&
\ee
which can also be obtained from the algebraic relations for $\frak{osp}(1|4)$ in the limit
$l_r\to 0$.
These relations together with the pure bosonic ones present the supersymmetric
extension of the second realization of
Poincar\'e algebra. For brevity, we call it the second Poincar\'e
superalgebra.  Obviously, the invariant length parameter $l$ appears
in the fermionic operators as well as in the bosonic operators in
the second Poincar\'e superalgebra. The second Poincar\'e
superalgebra have dramatically different meaning from the ordinary
Poincar\'e superalgebra, though their algebraic structures are the
same.

\subsection{Supersymmetric extension of anti-Newton-Hooke algebra}

It has been shown \cite{HTWXZ3} that the algebra $\frak{n}_-$ and
$\frak{n}_{-2}$ can be acquired by the contraction from $ \frak {d}
_-$,
\be {\frak{n_-}}:&& H^-=\lim_{\substack{c_r,\,l_r\to \infty \\
\nu_r=\nu {\rm\;fixed}}}H_r^-,
  \quad \vect{P}=\lim_{\substack{c_r,\,l_r\to \infty \\ \nu_r=\nu}} \vect{P}_r^-,
  \quad \ \ \vect{K}^{\frak g}=\lim_{\substack{c_r,\,l_r\to \infty \\
  \nu_r=\nu}}\vect{K}_r,\ \ \, \vect{J},\\
  \mbox{and\quad}{\frak n}_{-2}:&&
  H^-= \lim_{\substack{c_r,\,l_r \to 0 \\ \nu_r=\nu}} H_r^-,\ \quad \,
  \vect{P}'= \lim_{\substack{c_r,\,l_r \to 0 \\ \nu_r=\nu}} \d
  {l_r^2}{l^2}\vect{P}_r^-,\quad \, \vect{K}^{\frak c}=
  \lim_{\substack{c_r,\,l_r \to 0 \\ \nu_r=\nu}} \d
  {c_r^2}{c^2}\vect{K}_r,\ \vect{J}. \quad %
\ee
The supersymmetric extensions of the generators for $\frak{n}_-$ are
\be
H^-=\r_t+\frac {c^2}{l^2}tx^\mu\r_{x^\mu},\quad P_i=\r_{x^i},\quad \mbox{and} \quad
 K^{\frak g}_i=t\r_{x^i}  \label{n-ssext}
\ee
together with Eq.(\ref{SAM}).
The fermionic generator for the anti-Newton-Hooke
superalgebra is attained in the limit of $c_r,l_r \to \infty$ but
$\nu_r=\nu$,
\be Q^{\frak{n}_{-}}&:=&\lim_{\substack{l_r, c_r\to \infty\\
\nu_r={\rm fixed} }} Q_r = \La^{\frak {n}_-}\left (i\frac \r {\r\bar\th}
-\frac 1 2\sqrt{\si_{\frak n}}(\ga^i\th)\frac \r {\r x^i}\right ) \\ %
\mbox{with \qquad }
\La^{\frak {n}_-} &=&\left (\frac{1+\sqrt{\si_{\frak n}}}{2\sqrt{\si_{\frak n}}}
\right)^{\frac 1 2}\left (I+\frac {i\nu t \ga _0}{1+\sqrt {\si_{\frak n}}}\right ), %
\ee
where $ \si_{\frak {n}}=1+\nu^2 t^2\ $.  With these generators,
the $AdS$ superalgebra contracts to the $\frak{n}_-$ superalgebra,
\be
\{Q^{\frak{n}_-}, \bar {Q}^{\frak{n}_-}\} =i\ga^i
P_i-2\nu\Si^{0i}K_i^{\frak g},
  &\quad \quad & \\
\,  [J_{ij}, Q^{\frak{n}_-}]=-\Si_{ij} Q^{\frak{n}_-},
\qquad\qquad &\quad  \quad &
  [K_i^{\frak{g}}, Q^{\frak{n}_-}]=0,\\
\,  [H^-, Q^{\frak{n}_-}]=\d
  {i\nu}{2}\ga_0Q^{\frak{n}_-},
\qquad\quad \ \ &\quad  \quad &   [P_i,Q^{\frak{n}_-}]=0,
  \ee
together the commutators for the bosonic generators as shown in Table I.

In the process of the limit $c_r,l_r \to 0$, $\nu_r=\nu$,
$\si \to 1+\nu^2 t^2 -l_r^{-2}\dl_{ij}x^ix^j$ cannot
preserve positive.  Therefore, in the limit the fermionic generators are ill-defined
even though the bosonic generators might be written down in the same way as
for the supersymmetric extenssion of $\frak{n}_-$.  In other words, no supersymmetric
extension of the algebra $\frak{n}_{-2}$  can be obtained from the contraction of
$\frak{osp}(1|4)$.

\subsection{Supersymmetric extension of $\frak{h}_-$ and $\frak{p}'$}

The generators of $HN_-$ algebra ($\frak{h}_-$) and para-Poincar\'e
algebra ($\frak{p}'$) can be contracted from $AdS$ algebra as
  follows \cite{HTWXZ3}:
  \be
  {\frak h}_-: && \quad H=\lim_{\substack{c_r \to 0}}H_r^-,\qquad \vect{P}^-,\quad
  \vect{K}^{\frak{c}}=\lim_{\substack{c_r \to 0}} \d {{c_r}^2} {c^2}
  \vect{K}_r,\quad \vect{J},\\
   \mbox{and\quad}{\frak p}':&& \quad
   H'=\lim_{\substack{c_r \to \infty}} \d {c^2}{c_r^2}H_r^-,\ \,\vect{P}^-,\quad
   \vect{K}^{\frak{g}}=\lim_{\substack{c_r \to \infty}}\vect{K}_r,\qquad
   \vect{J}.
\ee
Their supersymmetric extensions are
\be
 {\frak h}_-: &&H=\r_t,\qquad  \qquad {P^{\frak{h}}}_i^-=\r_{x^i}+l^{-2}x_ix^\mu\r_{x^\mu}
 +\frac 1 {l^2} \frac {x^j}{1+\sqrt{\si_b}}\bar \th \Si_{ji}\frac{\r}{\r \bar \th},\quad
 K_i^{\frak{c}}=-\frac 1 {c^2}x_i\r_t,\\
{\frak p}':&&
 H'=\frac {c^2}{l^2}tx^\mu\r_{x^\mu},\ \, {P^{\frak{p}'}}_i^-=\r_{x^i}+l^{-2}x_ix^\mu\r_{x^\mu}
  \pm \frac 1 {l^2} \bar \th \Si_{0i}\frac{\r}{\r \bar \th},\qquad \qquad \ \
 K_i^{\frak{g}}=t\r_{x^i},
\ee
and Eq.(\ref{SAM}), where $\si_b=\displaystyle\lim_{c_r\to 0} =1-l^{-2}\dl_{ij}x^i x^j >0$
gives the
the domain condition when $c_r \to 0$.  The condition defines the domain inside a ball of radius
$l$ and thus the $\si$ in the limit is denoted with a subscript $b$.  The third term in
${P^{{\frak p}'}}_i^-$ is taken the same sign as $t$.  It is remarkable that
 the spatial translation
generators for $\frak{h}_-$ and $\frak{p}'$ are the
same before supersymmetric extension, but after that they become different.

In the $c_r\to 0$ limit, the $AdS$ superalgebra with scaled $\th$
contracts to $ \frak h _-$ superalgebra
if $\epsilon \to 0$ as $\varepsilon_b:=\sqrt{{c_r}/{c}}\to 0$ and if the fermionic
generator is chosen as
\be %
Q^{\frak{h}_-}&:=&\lim_{c_r \to 0}\varepsilon_b Q_r= \La^{\frak h_-}\left (
i\frac \r {\r \bar \th} -\frac{\sqrt {\si_{\frak b}}}{2c}\ga_0\th\frac \r {\r t}\right ), \\
\mbox{with\qquad}  \La^{\frak h_-}&=&\left (\frac
{1+\sqrt{\si_b}}{\sqrt{\si_b}}\right )^{\frac 1 2}\left (I+\frac 1 l \frac{ix^i
\ga_i}{1+\sqrt{\si_b}}\right).
 \ee
The algebraic relations involving the fermionic generators contract to
 \be
\,[J_{ij},Q^{\frak{h}_-}]=-\Si_{ij}Q^{\frak{h}_-},\qquad \qquad \ &&
  [K_i^{\frak{c}},Q^{\frak{h}_-}]=0,\\
\   [H,Q^{\frak{h}_-}]=0,
\qquad \qquad\qquad \qquad && [P_i^-,Q^{\frak{h}_-}]=\d{i}{2l}\ga_i
 \bar{Q}^{\frak{h}_-},\\
  \{Q^{\frak{h}_-},\bar Q^{\frak{h}_-}\}=i\ga_0 \frac H c -2\nu
  \Si^{0i}K_i^c.
\ee

On the other hand, when $c_r \to \infty$, the domain condition
$\si > 0 \to (c_r/c)^2 \nu^2 t^2>0$, which is always valid.
The $AdS$ superalgebra with scaled $\th$ contracts to $\frak p'$ superalgebra
if $\epsilon \to 0$ as $\varepsilon_c:=\sqrt{c/{c_r}} \to 0$ and if
the fermionic generator takes the form
\be %
Q^{\frak{p}'}&:=&\lim_{c_r \to \infty}\varepsilon_cQ_r  =
\La^{\frak{p}'}(i\frac \r {\r\bar \th}-\frac {\nu t} {2l}\ga ^0\th
x^\mu \r_\mu
\mp\frac {\nu t}{2} \ga^i\th \frac \r {\r x^i}), \\
\mbox{with \qquad } \La^{\frak{p}'}&=&\frac {I \pm i\ga_0}{\sqrt 2}. \label{Lambda_p'}%
\ee
The sign in $Q^{\frak{p}'}$ is opposite to the sign of $t$.
The fermionic parts of algebraic relations of $\frak p'$ superalgebra are %
\be %
\,[J_{ij},Q^{\frak{p}'}]=-\Si_{ij}Q^{\frak{p}'},\qquad  \qquad \quad  & & [K_i^{\frak{g}},Q^{\frak{p}'}]=0,\\
\, [H',Q^{\frak{p}'}] =0,\qquad \qquad \qquad \quad \quad &&
   [P_i^-,Q^{\frak{p}'}]=\d{i}{2l}\ga_i{Q}^{\frak p'}, \\ %
\{Q^{\frak{p}'},\bar Q^{\frak{p}'}\}
   =i\ga_0 \frac {H'} c -2\nu \Si^{0i}K_i^{\frak g}.&&
\ee

\subsection{Supersymmetric extension of Galilei algebra}

The generators of Galilei algebra ($\frak g$) can be contracted from \AdS\ algebra as
follows \cite{HTWXZ3}:%
\be %
  H=\lim_{\substack{c_r,l_r\to \infty \\ \nu_r \to 0  }}H_r^-,
   \qquad \  \vect{P}=\lim_{\substack{c_r,l_r\to \infty  \\ \nu_r \to 0 }}\vect{P}_r^-,\
   \qquad \vect{K}^{\frak g}=\lim_{\substack{c_r,l_r\to \infty \\ \nu_r \to 0}}
   \vect{K}_r,\qquad \ \vect{J}.
\ee
The supersymmetric extensions of $H$, $\vect{P}$ and $\vect{K}$ are the same as the
generators before supersymmetric extension.
The fermionic generator of supersymmetric extension of $\frak g$ is
\be \label{GalileiQ} %
Q^{\frak{g}} &:=&\lim_{\substack{c_r,l_r\to \infty \\ \nu_r \to
0}}Q_r =\lim_{c_r\to \infty}Q^{\frak{p}}=\lim_{\nu_r\to
0}Q^{\frak{n}_-} = i\frac \r{\r \bar \th}-\frac {\ga^i \th}{2} \frac
\r {\r x^i}.
 \label{GalileitQ}%
\ee
They span the Galilei superalgebra.
In particular,
 \be
 \,\  [J_{ij},Q^{\frak g}]=-\Si_{ij}Q^{\frak g},
\quad &\quad \quad &
  [K_i^{\frak{g}},Q^{\frak g}]=0,\\
 \, [H,Q^{\frak g}]=0, \qquad \quad \quad
 &\quad \quad & [P_i,Q^{\frak g}]=0,\\
 \{Q^{\frak g},\bar{Q}^{\frak g}\}=i \ga^i
  P_i. \quad \ \ &&
  \ee

In the limit of $c_r, l_r \to 0$, $\nu_r \to \infty$, $AdS$ algebra ($\frak{d}_-$)
contracts to the second Galilei algebra ($\frak g_2$).  But, since $\sqrt{\si}$
cannot remain real in the limiting process, no supersymmetric extension of
$\frak{g}_2$  can be obtained in this way.

\subsection{Supersymmetric extension of $\frak c$ and $\frak{c}_2$}

As the contraction from $\frak d_-$, the generators of Carroll
algebra $\frak c$ are \cite{HTWXZ3}:
\be %
  {\frak c}:&& H=\lim_{\substack{l_r \to \infty \\ c_r \to
  0}}H_r^-,\qquad \quad \vect P=\lim_{\substack{l_r \to \infty \\ c_r \to
  0}}\vect{P_r^-},\qquad \ \vect{K}^{\frak c}=\lim_{\substack{l_r \to \infty \\ c_r \to
  0}} \d {c_r^2} {c^2} \vect K_r,\quad  \vect J,\\
  \mbox{and\quad}{\frak c}_{2}:&&
  H'=\lim_{\substack{l_r \to 0 \\ c_r \to\infty}}
  \d{\nu^2}{\nu_r^2}H_r^-,\ \quad \vect P'=\lim_{\substack{l_r \to 0 \\ c_r
  \to\infty}}\d{l_r^2}{l^2}\vect P_r^-,\ \quad \vect K^{\frak g}=\lim_{\substack{l_r \to 0 \\ c_r
  \to\infty}}\vect K_r,\qquad \vect J. %
\ee
The expressions for $H$, $H'$, $\vect{P}$, $\vect{P}'$, $\vect{K}^{\frak{c}}$
and $\vect{K}^{\frak{g}}$ after supersymmetric extension remain their forms before
supersymmetric extension.

In the limit of $l_r\to \infty$ and $c_r\to 0$, the fermionic
generators tend to %
\be %
Q^{\frak{c}} &:=&\lim_{\substack{l_r\to
\infty \\ c_r\to 0}}\varepsilon_b Q_r =\lim_{c_r\to 0}\varepsilon_b
Q^{\frak{p}}_r=\lim_{l_r\to \infty }Q^{\frak{h}_-}_r =i\frac \r {\r
\bar \th}-\frac {\ga^0 \th}{2c}\frac \r{\r t} %
\ee
and the \AdS\ and Poincar\'e superalgebras with scaled $\th$ ($\epsilon = \varepsilon_b$),
and the anti-Hooke-Newton superalgebra contract to Carroll superalgebra,
\be   %
\{Q^{\frak c},\bar{Q}^{\frak c}\}=\frac {i} c \ga^0H, \qquad \ &&\\
\,  [J_{ij},Q^{\frak c}]=-\Si_{ij}Q^{\frak c},\qquad &&%
  [K_i^{\frak{c}},Q^{\frak c}]=0,\\
\,  [H,Q^{\frak c}]=0, \qquad \qquad \quad &&
[P_i,Q^{\frak c}]=0. %
\ee

On the other hand, if the following scale transformations are made,
\be
\th \to 
\varepsilon_c\varepsilon_m^2 \th = \sqrt{\frac{\nu l_r}{\nu_r l}}\th
\ee
in \AdS\ superalgebra,
\be \label{th4} %
\th\to 
\varepsilon_c \th 
\ee
in the second Poincar\'e superalgebra, or
\be %
\th\to \varepsilon_m^2 \th 
\ee
in para-Poincar\'e superalgebra, in the limit of $c_r\to \infty$ and
$l_r\to 0$ the fermionic generators will tend to %
\be %
Q^{\frak{c}_2} &:=&\lim_{\substack{l_r\to 0 \\ c_r\to \infty}}\varepsilon_c
\varepsilon_m^2Q =\lim_{c_r\to \infty}\varepsilon_cQ^{\frak{p}_2}_r
=\lim_{l_r\to 0}\varepsilon_m^2 Q^{\frak{p}'}_r = \La^{\frak{p}'} (i
\frac \r {\r \bar \th}-\frac {\nu t}{2l}\ga^0 \th x^\mu\r_\mu),
 \ee
where $\La^{\frak{p}'}$ is given by Eq.(\ref{Lambda_p'}),
and the corresponding superalgebras contract
to the second Carroll superalgebra,
\be  %
\{Q^{\frak c_2},\bar{Q}^{\frak c_2}\}=\frac {i\ga^0} c H', \quad &&\\
\,   [J_{ij},Q^{\frak c_2}]=-\Si_{ij} Q^{\frak c_2},
   &\qquad& [K_i^{\frak{g}},Q^{\frak c_2}]=0,\\
   \left[H',Q^{\frak c_2}\right]=0,\qquad \quad  &\qquad& [P'_i,Q^{\frak c_2}]=0.
\ee

\subsection{Supersymmetric extension of para-Galilei algebra}

   The generators of para-Galilei algebra $\frak g'$ can be
deduced from the contraction from $\frak{d} _-$ \cite{HTWXZ3}.
\be %
{\frak g'}:&& H'=\lim_{\substack{l_r,c_r \to\infty \\ \nu_r \to
\infty}} \d{\nu^2}{\nu_r^2}H_r^-,\ \ \vect P
=\lim_{\substack{l_r,c_r \to\infty \\ \nu_r \to\infty}}\vect P_r^-,
\qquad \ \ \vect K^{\frak g}=\lim_{\substack{l_r,c_r \to\infty \\
\nu_r \to\infty}}\vect K_r,
  \  \quad \vect J .
\ee
Again, after the supersymmetric extension, $H'$, $\vect{P}$ and $\vect{K}^{\frak g}$
have the same expressions as those before supersymmetric extension.  And,
suppose that the scale factor $\varepsilon =\varepsilon_\nu:=\sqrt{\nu/\nu_r} $
in \AdS\ and anti-Newton-Hooke superalgebras, $\varepsilon=\varepsilon_m $
in para-Poincar\'e superalgebras.  In the limits of
$l_r, c_r, \nu_r \to \infty$, the fermionic generators for
$\frak{g}'$ superalgebra and the algebraic relations are attained,
\be %
Q^{\frak{g}'} &:=&\lim_{c_r,l_r,\nu_r\to \infty} \sqrt{\d {\nu}
{\nu_r}}Q_r=\lim_{\nu_r\to \infty} \sqrt{\frac {\nu}
{\nu_r}}Q_r^{\frak{n}_-}
=\lim_{l_r\to \infty} \sqrt{\d {l_r} {l}}Q_r^{\frak{p}'} \nno \\
&=&\La^{\frak{p}'}(i\frac \r{\r\bar\th}
 \mp\frac {\nu t}2 \ga^i \th
\frac \r {\r x^i}),
\ee
and
\be %
\{Q^{\frak g'},\bar{Q}^{\frak g'}\}=-2\nu \Si^{0i}K_i^{\frak g}, \\
\,  [J_{ij},Q^{\frak g'}]=-\Si_{ij}Q^{\frak g'}, \qquad
&\quad \quad &
  [K_i^{\frak{g}},Q^{\frak g'}]=0,\\
\,  [H',Q^{\frak g'}]=0,  \qquad \quad \qquad 
 &\quad
 \quad & [P_i,Q^{\frak g'}]=0.%
\ee
During the limit of $l_r, c_r\to 0$ but $c_r/l_r^2=c/l^2$ is taken,
$\si$ cannot preserve positive for all $x^\mu$.
Therefore, the second para-Galilei algebra has no supersymmetric
extension.

\subsection{Summary}

The femionic generators in Beltrami realizations are gathered
together,
\be \nonumber %
&Q &=  (1-\frac 1 {4l}\bar\th \th)\La\{i(1+\frac 1 {4l}
\bar\th \th)\frac \r {\r \bar\th}+\frac i l\th\bar\th \frac \r {\r \bar
\th} -\frac i {4l}(\ga^\nu \th)\bar\th(\ga_\nu-\frac {2i} l \frac
{x^\mu}{1+\sqrt
\si}\Si_{\mu\nu})\frac {\r}{\r\bar\th}\\
&& -\frac 1 2 \sqrt{\si}\ga^\nu\th (\frac \r {\r x^\nu }
+\frac {l^{-2}x_\nu x^\mu}{1+\sqrt{\si}} \r_\mu)\},\setcounter{section}{3}
\setcounter{equation}{7} \\
&Q^{\frak{p}}&=i \frac \r {\r
\bar\th}-\frac 1 2 \ga^\mu \th \frac \r {\r x_{\mu}}, \setcounter{section}{4}
\setcounter{equation}{4} \\
&Q^{\frak{p}_2}&=\La^{\frak{p}_2}(i\frac \r{\r\bar\th}-\frac 1 {2l^2}
(\ga^\nu \th) x_\nu x^\mu \r_\mu),  \setcounter{equation}{12}  \\
&Q^{\frak{n}_{-}}&=  \La^{\frak{n}_-}(i\frac \r {\r\bar\th}-\frac 1 2\sqrt{\si_{\frak n}}
(\ga^i\th)\frac \r {\r x^i}) , \setcounter{equation}{19} \\ %
&Q^{\frak{h}_-} &=  \La^{\frak h_-}(i\frac \r {\r \bar \th}
-\frac{\sqrt {\si_{\frak b}}}{2c}\ga_0\th\frac \r {\r t}), \setcounter{equation}{27} \\
&Q^{\frak{p}'} &= \La^{\frak{p}'}(i\frac \r {\r\bar \th}-\frac {\nu
t} {2l}\ga ^0\th x^\mu \r_\mu
-\frac {\nu t}{2} \ga^i\th \frac \r {\r x^i}), \setcounter{equation}{33} \\
&Q^{\frak{g}} &=i\frac \r{\r \bar \th}-\frac {\ga^i \th}{2} \frac
\r {\r x^i}, \setcounter{equation}{40} \\
&Q^{\frak{c}} &=  i\frac \r {\r
\bar \th}-\frac {\ga^0 \th}{2c}\frac \r{\r t} \setcounter{equation}{51} \\
&Q^{\frak{c}_2} &=  \La^{\frak{p}'} (i
\frac \r {\r \bar \th}-\frac {\nu t}{2l}\ga^0 \th x^\mu\r_\mu), \setcounter{equation}{58} \\
&Q^{\frak{g}'} &= \La^{\frak{p}'}(i\frac \r{\r\bar\th}
 \mp\frac {\nu t}2
\ga^i \th \frac \r {\r x^i}).\setcounter{equation}{64} %
\ee
%
They are spinor representation of $\frak{so}(3)$ algebra, satisfying %
\setcounter{equation}{59}
\be %
[J_{ij}, {\cal Q}] = - \Si_{ij}{\cal Q}. %
\ee
All realizations of the kinematical superalgebras satisfy the Jacobi
identities.  In addition, under the parity and time-reversal transformations
the fermionic coordinates $\th$ and $\bar\th$
transform as
\be
\Pi^{-1}\theta \Pi&=&i\ga^0 \theta,\qquad \Pi^{-1}\bar \theta \Pi=-i\bar\theta\ga^0 ,
 \\
\Theta^{-1}\theta \Theta &=&i\ga^1\ga^3 \theta,\quad
\Theta^{-1}\bar \theta \Theta=i\bar\theta\ga^1\ga^3 ,
\ee
and the fermionic generators $Q$ and  $\bar Q$ transform as
\be
\Pi^{-1}{\cal Q} \Pi&=&i\ga^0 {\cal Q}, \qquad \quad  \Pi^{-1}\bar {\cal Q} \Pi=-i\bar{\cal Q}\ga^0,
 \\
\Theta^{-1}{\cal Q} \Theta&=&-i\ga^1\ga^3 {\cal Q}, \quad  \Theta^{-1}\bar {\cal Q} \Theta
=-i\bar {\cal Q}\ga^1\ga^3 ,
\ee
respectively.
 They satisfy
\[
\Pi^{-1}\Pi^{-1}\theta \Pi\Pi =- \theta, \qquad \Theta^{-1}\Theta^{-1}\theta \Theta\Theta = -\theta
\qquad \mbox{and so on.}
\]
\omits{The parity behavior is the same as that in \cite{RembielinskiTybor},
but the time-reversal transformation law for the fermionic generator is
different from that in \cite{RembielinskiTybor}.}

In the contraction approach, the Beltrami time $t$ and Beltrami
spatial coordinates $x^i$ always keeps unchanged, while the
fermionic coordinates $\th$ and $\bar \th$  may change.

\begin{table}[th]
\caption{\quad Supersymmetric extension of \AdS\ algebra and its
contractions} {\small
\begin{tabular}{|c c c c c c c c|} %
\hline 
Symbol & $\begin{array}{c}\mbox{Fermionic}\\\mbox{generators}\end{array}$ %
& $[\cal H,{\cal Q}]$ \quad & $[\vect{\cal P},{\cal Q}]$ \quad      %
& $[\vect{\cal K},{\cal Q}]$\quad &$\{{\cal Q}, %
  \bar{\cal Q}\}$ & $\begin{array}{c} \mbox{scale} \\ \mbox{of }\th\end{array}$
  & Limit${}^{a}$ \\ 
\hline 
$\mathfrak{d}_-$ & $(Q,\bar Q)$ & $\frac {\nu} {2} \ga_{0 } Q$ %
& $ \frac {1} {2l}\ga_{i}{Q} $ & $- \frac{1}{c} \Si_{0i}Q$ %
&  $ i \ga^\mu P_\mu^- -\frac 1 l \Si^{\mu\nu}J_{\mu\nu}$ & 1 &
\\ 
$\mathfrak{p}$ & $(Q^{\frak p}, \bar Q^{\frak p}) $ %
& 0  & 0 &  $-\frac 1 c \Si_{0i} Q^{\frak p} $  & $ i\ga^{\mu}P_\mu $ %
& 1 & $\begin{array}{c}l_r\to\infty 
\smallskip \end{array}$ \\ 
$\mathfrak{p}_2$ & $(Q^{\frak{p}_2},\bar Q^{\frak{p}_2})$ %
& 0  & 0 & $-\frac 1 c \Si_{0i} Q^{{\frak p}_2}$ & $ i\ga^{\mu}P'_\mu$ %
& $\varepsilon_m^{2}$& $\begin{array}{c} l_r\to 0 
\end{array}$ \\
\hline 
$\mathfrak{n}_-$ & $(Q^{\frak{n}_-}, \bar Q^{\frak{n}_-} )$ %
& $\frac \nu 2 \ga_0  Q^{{\frak n}_-} $ & 0  & 0   %
& $ i\ga^i P_i-2\nu \Si^{0i}K_i^{\frak g}$ & 1 &
$\begin{array}{c }l_r, c_r\to\infty,\\ \nu_r=\nu 
\smallskip \end{array}$ \\ 
$\mathfrak{h_-}$ &$(Q^{\frak{h}_-}, \bar Q^{\frak{h}_-})$ & 0 %
& $ \frac 1 {2l}\ga_i  Q^{{\frak h}_-} $ &0  %
& $ \frac {i \ga^0} {c} H-2\nu \Si^{0i}K_i^{\frak c} $ & $\varepsilon_b$
&$\begin{array}{c}c_r\to 0 
\smallskip \end{array}$ \\ 
$ \mathfrak{p}'$ %
&$(Q^{\frak{p}'},\bar Q^{\frak{p}'} )$ & 0 %
& $\frac 1 {2l}\ga_i  Q^{{\frak p}'} $ &0 %
& $ \frac {i \ga^0} {c} H'-2\nu \Si^{0i}K_i^{\frak g}$ %
& $\varepsilon_c$ & $\begin{array}{c} c_r\to \infty 
\smallskip \end{array}$ \\ 
$\mathfrak{g}$ & $( Q^{\frak g},\bar Q^{\frak g})$& 0 & 0 %
&  0    & $-i\ga^{i}P_i$ &1 & $\begin{array}{c}l_r, c_r\to\infty,\\ \nu_r\to 0
\smallskip\end{array}$ \\ 
$\mathfrak{c}$ & $( Q^{\frak c}, \bar Q^{\frak c} )$ %
& 0& 0& 0&  $ -\frac {i}{c}\ga^0 H$ & $\varepsilon_b$ &$\begin{array}{c}l_r\to\infty,\\
 c_r\to 0
 \smallskip \end{array}$\\ 
$ \mathfrak{c}_2$ & $(Q^{\frak{c}_2}_\al,\bar Q^{\frak{c}_2}_{\dot \al} )$ %
& 0& 0& 0 & $-\frac {i}{c}\ga^0 H' $ %
& $\varepsilon_\nu\varepsilon_m$&$\begin{array}{c} l_r\to 0,\\ c_r\to \infty
\smallskip\end{array}$\\ 
$\mathfrak{g}'$ & $(Q^{\frak{g}'}_\al, \bar Q^{\frak{g}'}_{\dot \al})$ %
& 0 & 0 & 0  &  $-2\nu \Si^{0i}K_i^{\frak g}$ %
& $\varepsilon_\nu$ &$\begin{array}{c}l_r, c_r\to\infty, \\\nu_r\to\infty
\end{array}$\\ 
\hline 
\end{tabular}}
{\footnotesize \begin{flushleft} $a$\quad \omits{$Q_r$ and $\th_r$
are the fermionic generator and fermionic coordinate,
of the \AdS\
superaglebra, respectively, when the invariant parameter(s) is variable. \\
$b$\qquad} $\varepsilon_l=\sqrt{l/l_r}, \varepsilon_m=\sqrt{l_r/l},
\varepsilon_b=\sqrt{c_r/c},
\varepsilon_c=\sqrt{c/c_r},\varepsilon_\mu=\sqrt{\nu_r/\nu},
\varepsilon_\nu=\sqrt{\nu/\nu_r}$.
\end{flushleft}}
\end{table}

The characteristic (anti-)communitators of all supersymmetric
extension of the \AdS\ algebra and its contractions are listed in
TABLE II. The relationship of these superalgebras can be seen more
clearly in FIG. 1.  The realizations of possible kinematical
algebras without supersymmetric extension are depicted in light
grey.
\begin{figure}[t]
\includegraphics[scale=0.6]{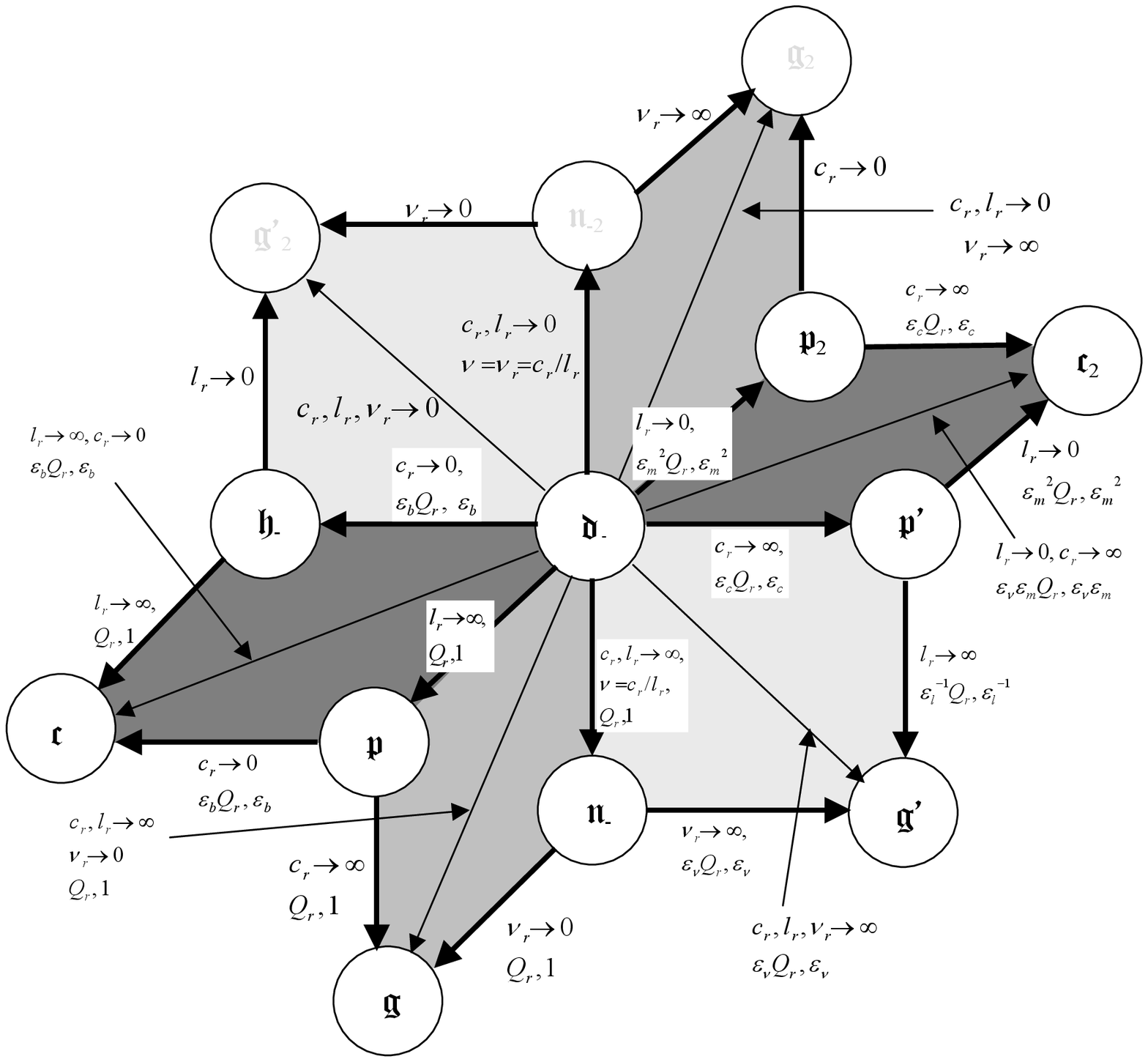}\\
\caption{Contraction scheme for the \AdS\ superalgebra and its
contractions. The second version of anti-Newton-Hooke, Galilei,
para-Galilei algebras have no supersymmetric extension. Thus,
$\frak{n}_{-2}$, $\frak{g}_2$ and $\frak{g}'_2$ are denoted in light
grey. }  \label{Fig:arn}
\end{figure}


\section{Concluding remarks}

It has been shown in  Ref. \cite{HTWXZ3} that if the realizations of
possible kinematical algebras are taken into account, more possible
kinematics possessing the same $\frak{so}$(3) isotropy generated by
$\vect{J}$ will be obtained than in the Bacry-L\'evy-Leblond's
classical work \cite{BLL}, having dramatically different physical
and geometrical meanings \cite{{Huang1},{HTWXZ1},{HTWXZ2},{HTWXZ3}}.
For example, the second realization of Poincar\'e group will not
preserve the Minkowksi metric.  In this paper, we apply the same
method as in Ref. \cite{HTWXZ3} to study the supersymmetric
extension of the \AdS\ algebra and present more explicit Beltrami
realizations of the \AdS\ superalgebra and its contractions. \omits{In
addition to the supersymmetric extensions of the $\frak{d}_-$,
$\frak{p}$,
 $\frak{n}_-$, $\frak{g}$ and $\frak{c}$ algebras in the literature \cite{CSdT},}
In particular,
the supersymmetric extensions of the $\frak{p}_2$, $\frak{h}_-$,
$\frak{p}'$, $\frak{c}_2$ and $\frak{g}'$ algebras are obtained.
\omits{Among these superalgebras, $\frak{h}_-$, $\frak{p}'$ and $\frak{g}'$
superalgebras are new ones.}  It is remarkable that the $\frak{p}_2$ and $\frak{c}_2$
superalgebras have the same algebraic structure as $\frak{p}$ and
$\frak{c}$ superalgebras but have very different physical and
geometrical meanings.

In this paper, we also show that not all Beltrami realizations of
the possible kinematical algebras contracted from the \AdS\ algebra
have supersymmetric extension. The $\frak{n}_{-2}$, $\frak{g}_2$ and
$\frak{g}'_2$ algebras are such a kind of algebras.  The immediate
causes are that the Beltrami fermionic generators of \AdS\
superalgebra contain $\si=1+l^{-2}\eta_{\mu\nu}x^\mu
x^\nu>0$ and that it cannot keep positive when the corresponding
limits are taken. In fact, the geometries invariant under the
transformations generated by $\frak{n}_{-2}$, $\frak{g}_2$ or
$\frak{g}'_2$ come from the contraction from the double-time de Sitter space \cite{HTWXZ3}%
\be %
ds^2=-\frac 1 {\si}\left (\eta_{\mu\nu}dx^\mu
dx^\nu - \frac {\eta_{\mu\la} \eta_{\nu\si}x^\la x^\si dx^\mu
dx^\nu}{l^2\si}\right) %
\ee
with $\si<0$, which is also invariant under the
transformations generated by $\frak{so}(2,3)$ but has the signature
$(+,+,-,-)$. In other words, there is no $\frak{n}_{-2}$-,
$\frak{g}_2$- or $\frak{g}'_2$-invariant geometries contracted from
the \AdS\ spacetime. Therefore, it is reasonable that the \AdS\
superalgebra cannot give rise to the supersymmetric extensions of
$\frak{n}_{-2}$, $\frak{g}_2$ and $\frak{g}'_2$ algebra by the
contraction approach.   The supersymmetric extension of $\frak{n}_{-2}$,
$\frak{g}_2$ and $\frak{g}'_2$ might be attained by the contraction
from the supersymmetric extension of $\frak{so}(2,3)$ realized on the double-time
de Sitter spacetime (i.e. $\si<0$).  But, in that case, $\frak{p}$ will fail to be
extended supersymmetrically.

Finally, since the central charge is not taken into account in the
present paper, the fermionic generators (e.g.(\ref{GalileitQ})) for $\frak{g}$) are
different from the expressions in the literature \cite{ClarkLove}.

\section*{Acknowledgments}
This work is supported by National Natural Science Foundation of
China under Grant No.  11275207.

    \end{document}